\documentclass[12pt,nohyper,notoc]{JHEP3}
\usepackage{amssymb}
\usepackage{amsmath}
\usepackage{epsfig}
\newcommand{\beq}{\begin{equation}}
\newcommand{\eeq}{\end{equation}}
\newcommand{\beqn}{\begin{eqnarray}}
\newcommand{\eeqn}{\end{eqnarray}}
\newcommand{\bea}[1]{\beq\begin{array}{#1}}
\newcommand{\eea}{\end{array}\eeq}
\newcommand{\eq}[1]{(\ref{#1})}

\newcommand{\tr}{\mathop{\rm Tr}}

\newcommand{\cL}{{\cal L}}

\newcommand{\cO}{{\cal O}}

\newcommand{\JHEP}[3]{{\em JHEP }{\bf #1} (#2) #3}

\title{Finite Temperature QCD with Two Flavors of
Non-perturbatively Improved Wilson Fermions}

\author{V.G.~Bornyakov,$^{abc}$ M.N.~Chernodub,$^{ac}$ H.~Ichie,$^{a}$
Y.~Koma,$^{d}$ Y.~Mori,$^{a}$ Y.~Nakamura,$^{a}$ M.I.~Polikarpov,$^{c}$
G.~Schierholz,$^{ef}$ A.A.~Slavnov,$^{g}$ H.~St\"uben,$^{h}$ T.~Suzuki,$^{a}$
P.V.~Uvarov$^{c}$ and A.I.~Veselov$^{c}$ \\
\llap{$^a$}Institute for Theoretical Physics,
Kanazawa University, Kanazawa 920-1192, Japan \\
\llap{$^b$}Institute for High Energy Physics IHEP, 142284 Protvino, Russia\\
\llap{$^c$}Institute of Theoretical and
Experimental Physics ITEP, 117259 Moscow, Russia \\
\llap{$^d$}Max-Planck-Institut f\"ur Physik, 80805 M\"unchen, Germany \\
\llap{$^e$}Deutsches Elektronen-Synchrotron DESY, 22603 Hamburg, Germany \\
\llap{$^f$} John von Neumann-Institut f\"ur Computing NIC,\\
Deutsches Elektronen-Synchrotron DESY, 15738 Zeuthen, Germany\\
\llap{$^g$}Steklov Mathematical Institute, 117333 Moscow,
Russia\\
\llap{$^h$} Konrad-Zuse-Zentrum f\"ur Informationstechnik Berlin ZIB,
14195 Berlin, Germany\\ \vspace*{-0.75cm}
\begin{center} -- {\rm DIK} Collaboration -- \end{center}  }

\preprint{DESY 04/003\\KANAZAWA/2003-10\\ITEP-LAT/2003-07}

\abstract{
We study  QCD with two flavors of non-perturbatively improved Wilson fermions
at finite temperature on the $16^3 \, 8$ lattice. We determine the transition
temperature at lattice spacings as
small as $a \sim 0.12$~fm, and
study string breaking below the finite temperature transition.
We find that the static potential can be fitted by a two-state ansatz,
including a string state and a two-meson state.
We investigate the role of Abelian monopoles at finite
temperature.
}


\begin{document}

\section{Introduction}
\label{one}

Recently, efforts have been made to determine the critical
temperature $T_c$ of the  finite temperature transition in full
QCD with  $N_f=2$ flavors of dynamical quarks. The Bielefeld group
employed improved staggered fermions and an improved gauge field
action~\cite{Karsch:2000kv}. The CP-PACS collaboration used
improved Wilson fermions with mean field improved clover
coefficient and an improved gauge field action
\cite{AliKhan:2000iz}. Both groups were able to estimate $T_c$ in
the chiral limit, and their values are in good agreement with each
other. Still, there are many sources of systematic uncertainties.
The main one is that the lattices used so far are rather coarse.
In this paper we perform simulations on $N_s^3\,N_t=16^3\,8$
lattices at lattice spacings $a$ much smaller than in previous
works~\cite{Karsch:2000kv,AliKhan:2000iz}. To further reduce
finite cut-off effects, we use non-perturbatively $O(a)$ improved
Wilson fermions. A small lattice spacing is particularly helpful
in determining the parameters of the static potential.

In the presence of dynamical quarks the flux tube formed between static
quark-antiquark pairs is expected to break at large distances. At zero
temperature $T$ the
search for string breaking, i.e. flattening of the
static potential, has not been successful, if the static quarks are created
by the Wilson loop. (See e.g. \cite{Bolder:2001un}).
At finite temperature string breaking has been observed at $T<T_c$, if Polyakov
loops are used instead to create the quarks.
It is important to know the static potential at finite temperature
for phenomenology~\cite{Karsch:1987pv}. In particular, it is needed to compute
the dissociation temperatures for heavy quarkonia.
We suggest a new ansatz and confront that ansatz with our numerical data.

The dynamics of the QCD vacuum, and color confinement in particular,
becomes more transparent in the maximally Abelian gauge
(MAG)~\cite{AbelianProjections,MaA}. In this gauge the relevant degrees
of freedom are color electric charges, color magnetic monopoles,
`photons' and `gluons'~\cite{Kronfeld:1987vd}. There is evidence
that the monopoles condense in the low temperature phase of the
theory~\cite{MaA,Shiba:1994db}, causing a dual Meissner effect, which
constricts the color electric field into flux tubes, in accord
with the dual superconductor picture
of confinement~\cite{DualSuperconductor}. Abelian
dominance~\cite{AbelianDominance} and the dynamics of monopoles have been
studied in detail at zero temperature in quenched~\cite{Chernodub:1997ay}
and unquenched~\cite{Bornyakov:2001nd,Bornyakov:2003vx} lattice simulations.
It turns out that in MAG the string tension is accounted for almost
entirely by the monopole part of the Abelian projected gauge
field~\cite{Stack:1994wm,shiba2,Bali:1996dm,Bornyakov:2001nd}.
Furthermore, in studies of SU(2) gauge theory at nonzero
temperature~\cite{Ejiri:1996sh} it has been found that at the phase transition
the Abelian Polyakov loop shows qualitatively the same behavior as the
non-Abstain one. In this paper we extend the investigation of
Abelian dominance to full QCD at nonzero temperature.

The paper is organized as follows. In Section 2 we present the details
of our simulation. Furthermore, we describe the gauge fixing algorithm and the
Abelian
projection. Section 3 is devoted to the determination of the transition
temperature, and in Section 4 our results for the heavy quark
potential are presented. In Section 5 we study the monopole density in the
vacuum, as well as the action density in the vicinity of the flux tube.
We demonstrate that the flux tube disappears at large quark-antiquark
separations. Finally, in Section 6 we conclude. Preliminary results of
this work have been reported in~\cite{Mori:2003pm}.

\section{Simulation details}

We consider $N_f=2$ flavors of degenerate quarks. We use the Wilson
gauge field action and non-perturbatively $O(a)$ improved
Wilson fermions~\cite{Sheikholeslami:1985ij}
\begin{equation}
S_F = S^{(0)}_F - \frac{\rm i}{2} \kappa\, g\,
c_{sw} a^5
\sum_s \bar{\psi}(s)\sigma_{\mu\nu}F_{\mu\nu}(s)\psi(s),
\label{action}
\end{equation}
where $S^{(0)}_F$ is the original Wilson action, $g$ is the gauge
coupling and $F_{\mu\nu}(x)$ is the field strength tensor. The
clover coefficient $c_{sw}$ is determined non-perturbatively. This
action has been used in simulations of full QCD at zero
temperature by the QCDSF and UKQCD
collaborations~\cite{Booth:2001qp,Allton:2002sk}, whose results we
use to fix the physical scale and the $m_{\pi} \slash m_{\rho}$
ratio. At finite temperature the same action was used before in
simulations on $N_t=4$ and 6 lattices at rather large quark masses
and lattice spacings~\cite{Edwards:1999mm}.

Non-perturbatively improved $N_f=2$ Wilson fermions should not be
employed below $\beta \equiv 6/g^2 = 5.2$. In fact, $c_{sw}$ is
known only for $\beta \geq 5.2$~\cite{Jansen:1998mx}. The
simulations are done on $16^3\,8$ lattices at two values of the
coupling constant, $\beta=5.2$ and 5.25, and nine different
$\kappa$ values each. The parameters are listed in Table~\ref{tab1}.
They are also shown in Fig.~\ref{fig:constant},
together with lines of constant $r_0/a$ and constant
$m_\pi/m_\rho$ obtained at $T=0$~\cite{Booth:2001qp}. Note that
the lines of constant $T$ run parallel to the lines of constant
$r_0/a$. To check the finite size effects we have also done
simulations on the $24^3\,8$ lattice at $\beta = 5.2$,
$\kappa=0.1343$.

The dynamical gauge field configurations are generated on the
Hitachi SR8000 at KEK (Tsukuba) and on the MVS~1000M at Joint
Supercomputer Center (Moscow), using a Hybrid Monte Carlo , while
the analysis is done on the NEC SX5 at RCNP (Osaka) and on the
PC-cluster at ITP (Kanazawa). Our present statistics is shown in
Table \ref{tab1}. The length of the trajectory was chosen to be
$\tau=0.25$. We use a  blocked jackknife method to compute the
statistical errors of the observables and a bootstrap method to
compute the errors of the fit parameters.
\FIGURE[t]{
\centerline{\includegraphics[angle=-00,scale=0.45,clip=false]{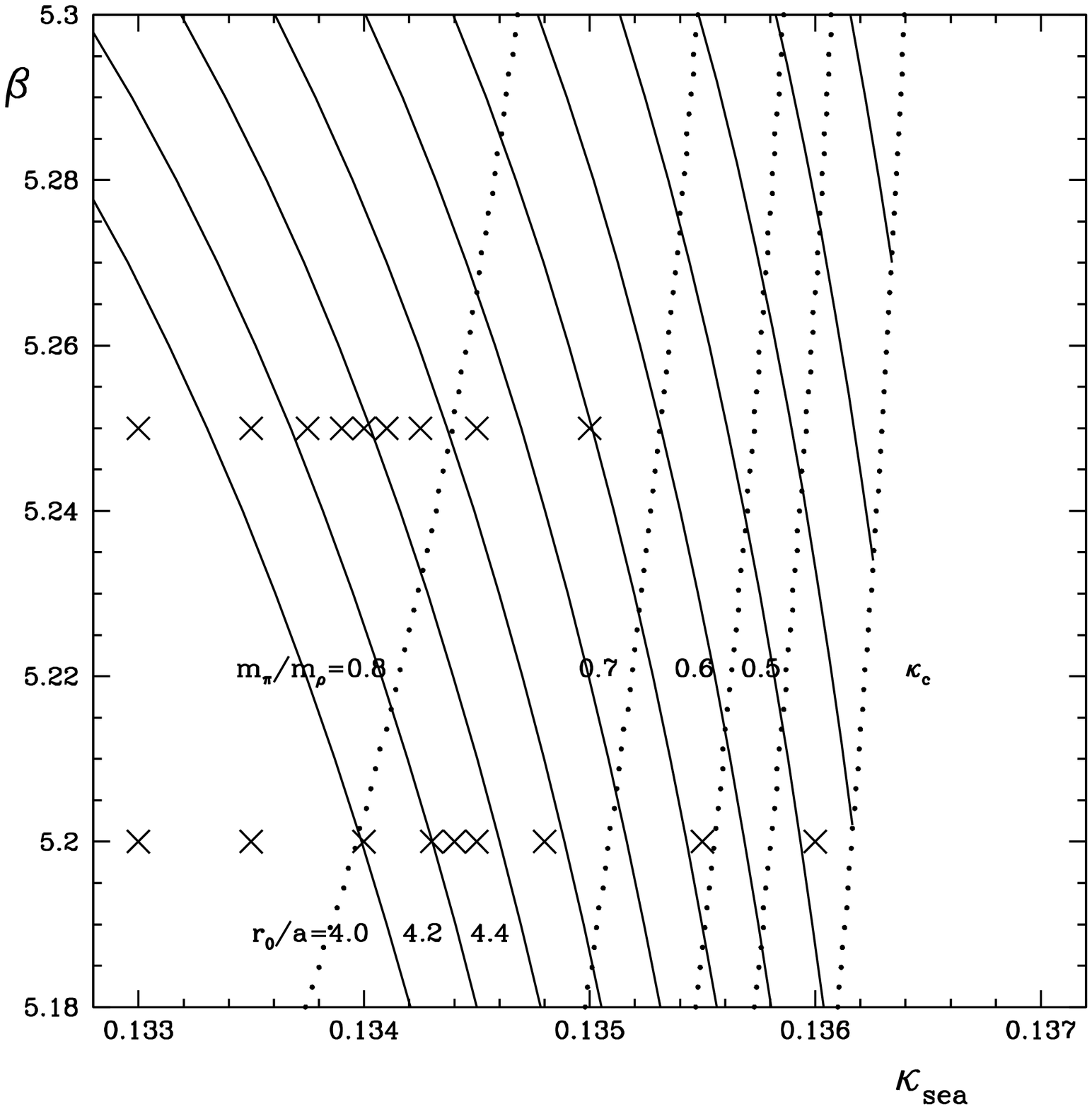}}
\vspace{-10mm}
\caption{Lines of constant $r_0/a$ (solid lines) and constant
$m_\pi/m_\rho$ (dotted lines) at $T=0$. Crosses indicate points
where simulations are done. }
\label{fig:constant}
\vspace*{0.5cm}
}
\TABLE[t]{
\begin{tabular}{|c|c|r||l|r|r|}
\hline
\multicolumn{3}{|c||}{$\beta=5.2$} & \multicolumn{3}{|c|}{$\beta=5.25$} \\
\hline
 $\kappa$ & Trajectories & \multicolumn{1}{c||}{$\tau_{int}$} &
\multicolumn{1}{c|}{$\kappa$} & \multicolumn{1}{c|}{Trajectories} &
\multicolumn{1}{c|}{$\tau_{int}$} \\
 \hline
 0.1330   & 7129  & 24      & 0.1330   & 1800~~~~  & 11    \\
 0.1335   & 4500  & 54      & 0.1335   & 7500~~~~  & 90    \\
 0.1340   & 3000  & 62      & 0.13375  & 9225~~~~  & 200   \\
 0.1343   & 6616  & 240     & 0.1339   & 12470~~~~ & 440   \\
 0.1344   & 8825  & 520     & 0.1340   & 19800~~~~ & 700   \\
 0.1345   & 6877  & 190     & 0.1341   & 14800~~~~ & 700  \\
 0.1348   & 5813  & 124     & 0.13425  & 5155~~~~  & 120   \\
 0.1355   & 5650  & 50      & 0.1345   & 2650~~~~  & 50    \\
 0.1360   & 3699  & 46      & 0.1350   & 1780~~~~  & 30    \\
 \hline
\end{tabular}
\caption{Parameters and statistics of the simulation, together with the
integrated autocorrelation time. The length of the trajectory is
$\tau = 0.25$.}
\label{tab1}
}
We compute the Polyakov loop
\begin{equation}
L(\vec{s}) = \frac{1}{3} \tr \prod_{s_4=1}^{N_t} U(s,4),
\end{equation}
$U(s,\mu)$ being the link variable, on every trajectory. {}From that we derive
the susceptibility
\begin{equation}
\chi = N_s^3 \sum_{\vec{s}}(\langle L^2(\vec{s})\rangle
- \langle L(\vec{s})\rangle^2), \label{susc}
\end{equation}
and the integrated autocorrelation time $\tau_{int}$. The
autocorrelation time is given in Table \ref{tab1} in units of trajectories.
Furthermore, we compute the Polyakov loop
correlator $\langle L(\vec{s}) L^{\dagger}(\vec{s}\,^\prime)\rangle$, from
which we obtain the static potential.
To reduce the error on the static potential,
we employ a hypercubic blocking of the gauge field as described in
\cite{Hasenfratz:2001hp}. We choose every $5^{th}$ to
$20^{th}$ trajectory, depending on the value of $\kappa$, to compute the
blocked Polyakov loop correlator.

We fix the MAG ~\cite{MaA} by maximizing the gauge fixing functional $F(U)$,
\begin{equation}
F(U) = \frac{1}{12V} \sum_{s,\mu} (|U_{11}(s,\mu)|^2 + |U_{22}(s,\mu)|^2 +
|U_{33}(s,\mu)|^2)
\end{equation}
with respect to local gauge transformations $g$ of the lattice gauge field,
\begin{equation}
U(s,\mu) \rightarrow U^g(s,\mu) = g(s)^\dagger U(s,\mu)  g(s+\hat{\mu})\,.
\end{equation}

To do so, we use the simulated annealing (SA) algorithm. The
advantage of this algorithm over the usual iterative procedure has
been demonstrated in \cite{Bali:1996dm} for the MAG and in
\cite{Bornyakov:2000ig} for the maximal center gauge in the SU(2)
gauge theory. In practice one does not find the global maximum of
the gauge fixing functional in a finite-time computation. For this
reason it has been proposed~\cite{Bali:1996dm} to apply the SA
algorithm to a number of randomly generated gauge copies and pick
that one with the largest value of $F$. By increasing the number
of gauge copies, one eventually reaches the situation, where the
statistical noise is larger than the deviation from the global
maximum. We use one gauge copy at $\beta=5.2$ and three gauge
copies at $\beta=5.25$. We have checked that by increasing the
number of gauge copies our results for the gauge dependent
quantities are left unchanged within the error bars.

To obtain Abelian observables, one needs to project the $SU(3)$ link matrices
onto the maximal Abelian subgroup $U(1) \times U(1)$ first. The original
construction~\cite{Brandstater:1991sn} is equivalent to
finding the Abelian gauge field $u(s,\mu) \in U(1) \times U(1)$ which
maximizes $|\mbox{Tr}\left(U(s,\mu)u^\dagger(s,\mu)\right)|^2$.
The Abelian counterpart of an observable is then obtained by
substituting
\beqn
u(s,\mu) = \mathrm{diag}(e^{{\rm i} \theta_1(s,\mu)},e^{{\rm i} \theta_2(s,\mu)},
e^{{\rm i} \theta_3(s,\mu)})\,,
\label{eq:UAbel}
\eeqn
for $U(s,\mu)$; $\sum_{i=1}^3 \theta_i(s,\mu) = 0$, so that $\det(u(s,\mu))=1$.
{}From eq.~(\ref{eq:UAbel}) we define plaquette angles \beqn
\theta_i(s,\mu\nu) =
\partial_{\mu} \theta_i(s,\nu)
- \partial_{\nu} \theta_i(s,\mu) \,,
\eeqn
where $\partial_\mu$ is the lattice forward derivative. The plaquette angles
can be decomposed into regular and singular components,
\beqn
\theta_i(s,\mu\nu) = \overline{\theta}_i(s,\mu\nu) + 2\pi m_i(s,\mu\nu)\,,
\eeqn
where $\overline{\theta}_i(s,\mu\nu) \in (-\pi;\pi]$, and $m_i(s,\mu\nu) \in
\mathbb{N}$ counts the number of Dirac strings piercing the given plaquette.
Note that $\sum_{i}\overline{\theta}_i(s,\mu\nu)=2\pi l$, $l=0,\pm 1$. If
$l=+1$ ($-1$) we substitute the largest (smallest)
$\overline{\theta}_i(s,\mu\nu)$ (of the three components) by
$\overline{\theta}_i(s,\mu\nu) - 2\pi$ ($+2\pi$), and similarly for
$m_i(s,\mu\nu)$, so that
$\sum_i\overline{\theta}_i(s,\mu\nu) = \sum_i m_i(s,\mu\nu) =0$.

The monopole currents, being located on the links of the dual lattice, are
defined by \cite{DGT}
\beqn
k_i(^*s,\mu) = \frac{1}{4\pi}\epsilon_{\mu\nu\rho\sigma} \partial_{\nu}
\overline{\theta}_i(s+\hat{\mu},\rho\sigma)
= -\frac{1}{2} \epsilon_{\mu\nu\rho\sigma} \partial_{\nu} 
m_i(s+\hat{\mu},\rho\sigma)\,.
\label{eq:mon:current}
\eeqn
They satisfy the constraint
\beqn
\sum_i k_i(^*s,\mu) = 0\,,
\label{eq:k-constraint}
\eeqn
for any ${}^* s,\mu$. The Abelian gauge fields $\theta_i(s,\mu)$
can in turn be decomposed into monopole (singular) and photon
(regular) parts: \beqn \theta_i(s,\mu) = \theta^{\rm mon}_i(s,\mu)
+ \theta^{\rm ph}_i(s,\mu)\,. \label{eq:decomposition} \eeqn The
monopole part is defined by~\cite{smit}:
\beqn
\theta^{\rm mon}_i(s,\mu) = -2 \pi \sum_{s'}  D(s-s') \, \partial_{\nu}'\,
m_i(s',\nu\mu)\,,
\label{eq:theta:mon}
\eeqn where
$\partial_{\nu}\,'$ is the backward lattice derivative, and $D(s)$
denotes the lattice Coulomb propagator.

The Abelian Polyakov loop is defined by
\beqn
L_{\rm Abel}(\vec{s}) = \frac{1}{3} \sum_{i=1}^3
L^{\rm Abel}_i(\vec{s})\,,
\qquad
L^{\rm Abel}_i(\vec{s}) = \exp \Bigl\{{\rm i} \sum_{s_4=1}^{N_t}
\theta_i(s,4) \Bigr\}\,.
\label{eq:W}
\eeqn
Similarly, we define the monopole and photon Polyakov loops~\cite{sibasuzuki}:
\begin{equation}
L_{\rm mon}(\vec{s}) = \frac{1}{3} \sum_{i=1}^3
L^{\rm mon}_i(\vec{s})\,,
\qquad
L^{\rm mon}_i(\vec{s}) = \exp \Bigl\{{\rm i} \sum_{s_4=1}^{N_t}
\theta_i^{\rm mon}(s,4) \Bigr\}\,,
\label{eq:monpl}
\end{equation}
\begin{equation}
L_{\rm ph}(\vec{s}) = \frac{1}{3} \sum_{i=1}^3 L^{\rm ph}_i(\vec{s})\,,
\qquad
L^{\rm ph}_i(\vec{s}) = \exp \Bigl\{{\rm i} \sum_{s_4=1}
\theta_i^{\rm ph}(s,4) \Bigr\}\,.
\label{eq:phpl}
\end{equation}

\section{Transition temperature}

The order parameter of the finite
temperature phase transition in quenched QCD is the Polyakov loop, and the
corresponding symmetry is global $Z(3)$. In the presence of dynamical
`chiral' fermions the chiral condensate
$\langle\bar{\psi}\psi\rangle$ is an order parameter (of the
chiral symmetry breaking transition). It is expected that there is no phase
transition at intermediate quark masses, only a crossover. Numerical
results show that both order parameters can be used to locate the transition
point at intermediate quark masses \cite{Karsch:2000kv}. We use
the Polyakov loop, because the calculation of the chiral condensate for Wilson
fermions requires renormalization and is rather involved.

It is instructive to plot the Polyakov loop in the complex plane as a function
of temperature, which has been done in Fig.~\ref{fig:ploop_dist}. We find
that the distribution is rather asymmetric, even at the lowest temperature,
favoring a positive value of ${\rm Re}\,L$. This is indeed what one
expects~\cite{Satz:2000hm}. The introduction
of dynamical quarks adds a term proportional to ${\rm Re}\,L$ to the effective
action, which results in a nonzero value of ${\langle L \rangle}$. The numbers
are given in Tables~\ref{tbl:polyakov:5.2} and \ref{tbl:polyakov:5.25}.
In the Tables we also give values for the Abelian and monopole Polyakov
loops separately.
\FIGURE[t]{
\centerline{\includegraphics[angle=-00,scale=0.45,clip=true]{Re-Im-monopole.eps}}
\vspace{-5mm}
\caption{Scatter plots of the Polyakov loop in the complex plane for
various temperatures.}
\label{fig:ploop_dist}
}
\TABLE[t]{
\hspace*{0.5cm}
\begin{tabular}{|c|l|l|l|l|c|}
\hline
$\kappa$ & \multicolumn{1}{c|}{${\langle L \rangle}$} &
\multicolumn{1}{c|}{${\langle L \rangle}_{\rm Abel}$}
& \multicolumn{1}{c|}{${\langle L \rangle}_{\rm mon}$} &
\multicolumn{1}{c|}{${\langle L \rangle}_{\rm ph}$} & $T/T_c$ \\
\hline
0.1330  & 0.0022(3)   &   0.014(2)  & 0.040(7)   &  0.2946(10) & 0.80  \\
0.1335  & 0.0027(7)   &   0.018(5)  & 0.054(16)  &  0.3007(9)  & 0.87  \\
0.1340  & 0.0034(5)   &   0.025(4)  & 0.079(14)  &  0.3052(7)  & 0.94  \\
0.1343  & 0.0092(13)  &   0.074(11)  & 0.235(35)  &  0.3113(6)  & 0.98  \\
0.1344  & 0.0131(18)  &   0.107(15) & 0.352(51)  &  0.3136(6)  & 1.00  \\
0.1345  & 0.0120(12)  &   0.098(10) & 0.319(33)  &  0.3145(5)  & 1.02 \\
0.1348  & 0.0207(11)  &   0.169(10) & 0.556(32)  &  0.3197(9)  & 1.06 \\
0.1355  & 0.0300(7)   &   0.235(5)  & 0.740(11)  &  0.3279(9)  & 1.18  \\
0.1360  & 0.0290(9)   &   0.236(6)  & 0.747(11)  &  0.3291(14) & 1.28  \\
\hline
\end{tabular}
\caption{The expectation values of the non-Abelian, Abelian, monopole and
photon Polyakov loops at $\beta=5.2$.}
\label{tbl:polyakov:5.2}
}

As we can see from Fig.~\ref{fig:constant}, increasing $\kappa$ at
a fixed value of $\beta$
increases the temperature $T \propto r_0/a$.
In Figs.~\ref{fig:ploop:a},~\ref{fig:ploop:b},~\ref{fig:ploop:c} we plot
the expectation values of the various Polyakov
loops of Tables~\ref{tbl:polyakov:5.2} and \ref{tbl:polyakov:5.25}
as a function of $\kappa$. While $\langle L\rangle$,
$\langle L_{\mathrm{Abel}}\rangle$ and $\langle L_{\mathrm{mon}}\rangle$
increase with increasing $\kappa$,
$\langle L_{\mathrm{ph}}\rangle$ stays approximately constant over the full
range of $\kappa$.
Furthermore, similar to the quenched theory, $L_{\mathrm{mon}}$ and
$L_{\mathrm{ph}}$ are virtually independent, which follows from
$\langle L_{\mathrm{Abel}}\rangle  \approx \langle L_{\mathrm{mon}}\rangle
\langle L_{\mathrm{ph}}\rangle$.
We have no explanation for the dip seen in $\langle L\rangle$ at
$\beta=5.25$ around $\kappa=0.1341$. We see some signal of metastability, but
we will need higher statistics to clarify this point.
A similar dip is seen on Fig.~\ref{fig:constant} of ref.~\cite{Edwards:1999mm},
where the same lattice action was used to study the phase
transition on small lattices.

\FIGURE[t]{
\centerline{\includegraphics[angle=-00,scale=0.45,clip=true]{polyakov.loop.nonabelian.eps}}
\vspace{-5mm}
\caption{The expectation value of the non-Abelian Polyakov loop
as a function of $\kappa$ for $\beta=5.2$ and 5.25.}
\label{fig:ploop:a}
}
\TABLE[t]{
\hspace{2cm}\begin{tabular}{|l|l|c|c|}
\hline
\multicolumn{1}{|c|}{$\kappa$} & \multicolumn{1}{c|}{${\langle L \rangle}$} &
${\langle L \rangle}_{\mathrm{mon}}$ & $T/T_c$ \\
\hline
0.1330 & 0.0025(7) & ~~0.060(13)~~ & ~~0.86~~ \\
0.1335   & 0.0076(12)   &    0.188(30) & 0.92\\
0.13375  & 0.0100(13)   &    0.273(35) & 0.96\\
0.1339   & 0.0118(11)   &    0.321(35) & 0.97\\
0.1340   & 0.0096(14)   &    0.248(40) & 0.99\\
0.1341   & 0.0086(17)   &    0.230(50) & 1.00\\
0.13425  & 0.0225(10)   &    0.604(22) & 1.02\\
0.1345   & 0.0255(9)   &       --      & 1.05\\
0.1350   & 0.0264(18)   &    0.706(25) & 1.12\\
\hline
\end{tabular}
\caption{The expectation values of the non-Abelian and monopole Polyakov
loops at $\beta=5.25$.}
\label{tbl:polyakov:5.25}}

\FIGURE[t]{
\centerline{\includegraphics[angle=-00,scale=0.45,clip=true]{polyakov.loop.abelian.1.eps}}
\vspace{-5mm}
\caption{The expectation value of monopole
and photon Polyakov loops as functions of
$\kappa$ for $\beta=5.2$ and 5.25.}
\label{fig:ploop:b}
}
\FIGURE[t]{
\centerline{\includegraphics[angle=-00,scale=0.45,clip=true]{polyakov.loop.abelian.2.eps}}
\vspace{-5mm}
\caption{The expectation value of the Abelian Polyakov loop as a function of
$\kappa$ for $\beta=5.2$. The product of monopole and photon
Polyakov loops is also shown.}
\label{fig:ploop:c}
}

The task is now to determine the transition temperature $T_c$. We
call the $\kappa$ value, at which the transition takes place,
$\kappa_t$. We identify $\kappa_t$ as the point, where the
Polyakov loop susceptibility (\ref{susc}) assumes its maximum. The
Abelian, monopole and photon Polyakov susceptibilities $\chi_{\rm
Abel}$, $\chi_{\rm mon}$ and $\chi_{\rm ph}$, respectively, are
defined similarly. The susceptibilities are given in Tables~\ref{tbl:sus:polyakov:5.2}
and \ref{tbl:sus:polyakov:5.25}, and they are plotted in
Figs.~\ref{fig:susceptibilities:a} and \ref{fig:susceptibilities:b}.
{}From the non-Abelian
susceptibility $\chi$ we find $\kappa_t=0.1344(1)$ at $\beta=5.2$
and $\kappa_t=0.1341(1)$ at $\beta=5.25$
where the central values and the errors 
are determined by the maxima of susceptibilities and
by the distances to neighbor data points, respectively
\FIGURE[t]{
\centerline{\includegraphics[angle=-00,scale=0.45,clip=true]{susceptibility.nonabelian.eps}}
\vspace{-5mm}
\caption{The non-Abelian Polyakov loop susceptibility as a function of
  $\kappa$ for $\beta=5.2$ and 5.25, respectively, together with the fit.}
\label{fig:susceptibilities:a}
}
\TABLE[t]{
\begin{tabular}{|c|c|c|c|c|}
\hline
$\kappa$ & $\chi $ & $\chi_{\mathrm{Abel}}$ & $\chi_{\mathrm{mon}}$ & $\chi_{\mathrm{ph}}$ \\
\hline
0.1330  & 0.072(3)    &  0.88(10)   &       7.7(8)     &   0.590(16)   \\
0.1335  & 0.094(5)    &  1.8(3)     &      17.4(2.4)    &   0.624(10)   \\
0.1340  & 0.095(12)   &  2.4(5)     &      25.5(4.9)    &   0.638(12)   \\
0.1343  & 0.115(17)   &  4.2(1.1)   &      46.1(12.)    &   0.653(10)   \\
0.1344  & 0.159(21)   &  7.7(1.5)   &      88.4(20.)    &   0.682(13)   \\
0.1345  & 0.151(29)   &  6.6(1.5)   &      70.4(14.)    &   0.671(16)   \\
0.1348  & 0.129(10)   &  5.7(8)     &      57.5(9.4)    &   0.705(25)   \\
0.1355  & 0.112(5)    &  2.7(4)     &      17.7(2.2)    &   0.686(27)   \\
0.1360  & 0.115(11)   &  2.3(5)     &      14.7(2.8)    &   0.734(29)   \\
\hline
\end{tabular}
\vspace*{0.5cm}
\label{tbl:sus:polyakov:5.2}
\caption{The same as in Table~\ref{tbl:polyakov:5.25} but for the susceptibility
of the Polyakov loop at $\beta=5.2$.} }
\FIGURE{
\centerline{\includegraphics[angle=-00,scale=0.45,clip=true]{susceptibility.abelian.eps}}
\vspace{-5mm}
\caption{The same as in Fig.~\ref{fig:susceptibilities:a} but for the Abelian,
  the monopole and the photon Polyakov loop susceptibilities.
  The Abelian (photon) Polyakov loop susceptibility has been enhanced by
  a factor of 10 (100).}
\label{fig:susceptibilities:b}
}
\TABLE[t]{
\hspace*{2cm}\begin{tabular}{|c|c|c|}
\hline
$\kappa$ & $\chi$ & $\chi_{\mathrm{mon}}$ \\
\hline
0.1330   &  0.07(1)  &  15(5)      \\
0.1335   &  0.11(4)  &  65(15)     \\
0.13375  &  0.14(4)  &  105(25)    \\
0.1339   &  0.15(4)  &  75(20)     \\
0.1340   &  0.18(5)  &  106(22)    \\
0.1341   &  0.24(7)  &  150(30)    \\
0.13425  &  0.13(3)  &  37(10)     \\
0.1345   &  0.10(3)  &  --          \\
0.1350   &  0.09(3)  &  53(18)     \\
\hline
\end{tabular}\hspace*{2cm}
\label{tbl:sus:polyakov:5.25}
\vspace*{0.25cm}
\caption{The same as in Table~\ref{tbl:sus:polyakov:5.2} but for
the susceptibility of the Polyakov loop at $\beta=5.25$.}
}
This translates into $T_c r_0 = 0.53(1)$ at $\beta=5.2$ and
$T_c r_0 = 0.56(1)$ at $\beta=5.25$, where the numbers of $r_0/a$ at
$\kappa_t$ have been obtained by interpolation of the
$T=0$ results~\cite{Booth:2001qp}.
Taking $r_0=0.5$ fm to fix the scale, we obtain in physical units
\beqn
T_c &=&210(3)~ {\rm MeV},\,\,\,  \beta=5.2 
\label{eq:crit} \\
T_c &=&219(3)~ {\rm MeV},\,\,\,  \beta=5.25\,. \nonumber 
\eeqn
By interpolating $m_{\pi}/m_{\rho}$, given in~\cite{Booth:2001qp}, to
$\kappa_t$ we obtain at the transition point $m_{\pi}/m_{\rho}=0.77$ at
$\beta=5.2$ and $0.81$ at $\beta=5.25$.
Similarly, we can compute the temperature $T$ at our various $\kappa$ values.
The result is given in the last column of Table~\ref{tbl:polyakov:5.2} and 
Table~\ref{tbl:polyakov:5.25}
in the form of $T/T_c$.

The susceptibilities $\chi_{\rm Abel}$ and  $\chi_{\rm mon}$  have
maxima at the same $\kappa$ value as the non-Abelian susceptibility.
We find that $\chi_{\rm Abel} \approx
\langle L_{\rm ph}\rangle^2 \chi_{\rm mon}$. This follows from our earlier
observation, namely that $L_{\rm mon}$ and $L_{\rm ph}$ are independent, and
the smallness of $\chi_{\mathrm{ph}}$.
The non-Abelian susceptibility is 10 to 50 times smaller than its
Abelian counterpart. The photon susceptibility does not show any change at
the critical temperature, as expected. We conclude, that the
monopole degrees of freedom
are most sensitive to the transition, as was the case in the quenched
theory.

In Fig.~\ref{fig:temp} we compare our results for $T_c$ with those
of Refs.~\cite{Karsch:2000kv} and \cite{Edwards:1999mm},
where we have assumed $\sqrt{\sigma}=425$~MeV.
Our results are in quantitative agreement with the results of the Bielefeld
group. This is reassuring, as \cite{Karsch:2000kv} and \cite{Edwards:1999mm}
work at larger lattice spacing.

\FIGURE[t]{
\centerline{\includegraphics[angle=-00,scale=0.45,clip=true]{critical.t.eps}}
\vspace{-8mm}
\caption{The transition temperature as a function of
$m_\pi/m_\rho$ from this work (triangles), and from
\cite{Karsch:2000kv} (squares) and \cite{Edwards:1999mm} (circle),
respectively.} \label{fig:temp}
}
\TABLE[t]{
\vspace{5mm}
\begin{tabular}{|c|c|c|c|c|}
\hline
$\beta$ & \multicolumn{3}{|c|}{$5.2$} & $5.25$ \\ \hline
$\chi$        & non-Abelian        & Abelian        & monopole & non-Abelian  \\ \hline
$\kappa_t$   & 0.13457(8)& 0.13459(5)& 0.13457(4) & 0.13407(2)  \\
$T_c$        & 215.7(2.5)& 216.3(1.0)& 215.7(1.2) & 217.9(5)   \\
$\nu$        & 0.11(4)   & 0.41(17)  & 0.48(14)   & 0.18(3)   \\
 \hline
\end{tabular}
\caption{The parameters of the fit \eq{eq:suscept}. The critical temperature
$T_c$,
determined from fits of various susceptibilities, is given in MeV.}
\label{tbl:sus}   }

We fit the susceptibility in the transition region by~\cite{Schiller}
\beqn
\chi^{\mathrm{fit}}(\kappa) = \frac{C_1}{{\Bigl(C_2 + (\kappa - \kappa_t)^2\Bigr)}^\nu}\,,
\label{eq:suscept}
\eeqn
where $C_{1,2}$, $\kappa_t$ and $\nu$ are taken as fit parameters. The
fit values, and the corresponding values for $T_c$, are presented in
Table~\ref{tbl:sus}.

The fits are shown by the dashed lines in Fig.~\ref{fig:susceptibilities:a}.
All fits gave $\chi^2/{\rm
dof} = O(1)$. 
We observe that the non-Abelian, Abelian and monopole susceptibilities
(available at $\beta=5.2$) give the  same $\kappa_t$ within error bars.
In principle, this fit should give the same value of $\kappa_t$
as our previous estimate by the maximum of susceptibility but with better 
precision. This is indeed the case for $\beta=5.25$. 
For $\beta=5.2$ the result is shifted to higher
$\kappa$. This shift is due to the data at $\kappa=0.1348$, which is
rather far from the transition. To be on a safe side we take
the values quoted in eq.(\ref{eq:crit}) as the critical temperature and respective 
errros for both values of $\beta$.

Because $N_s/N_t=2$ in our simulations, the question of finite volume
effects is essential.
To check for finite volume effects we have performed simulations
at $\beta=5.2$, $\kappa=0.1343$ on $24^3\times 10$ lattices. The
value of $\kappa$ was chosen close to the transition point, where
finite volume corrections are expected to be largest. We found
$\langle L \rangle = 0.0098(10)$ and $\chi = 0.099(13)$, as
compared to $\langle L \rangle =0.0092(13)$ and $\chi=0.115(17)$,
respectively, on the $N_s=16$ lattice. In both cases the numbers
agree within the error bars, so that we do not reckon with large
effects.

\section{Heavy quark potential}

\subsection{Ansatz}
\label{IVA}

One of the characteristic features of full QCD is breaking of the string
spanned between static quark and anti-quark pairs. String breaking will
manifest itself in a type of screening behavior of the heavy quark potential.
At zero temperature no clear evidence for string breaking has been found (in
QCD) yet. The reason is, so it is believed, that the Wilson loop
has very small overlap with the broken string state. The expectation value
of the Wilson loop for large distances $r$
can be written as
\begin{equation}
  \langle W(r,t) \rangle = C_V(r)\, e^{-( V_0+V_{string}(r) )\, t} +  C_E(r)\, e^{-2E_{sl}
  \cdot t} + ...\,,
\label{wl1}
\end{equation}
where $V_{\rm string}(r)$ is the usual confining potential,
$V_{\rm string}(r)=-\alpha/r +\sigma r$, $E_{\rm sl}$ is the
static-light meson energy, and $V_0$ is the self-energy. The latter
is divergent in the continuum limit $a \to 0$. The overlap with
the string state, $C_V(r)$, is of the order of one, while the
overlap with the broken string, $C_E(r)$, appears to be small. An
estimate~\cite{Aoki:1998sb} is: $C_E(r) \sim e^{-2m_{sl}\cdot r}$,
where $m_{sl} = E_{sl} - V_0/2$ is the so called binding energy of
the static-light meson or, in other words, the constituent quark
mass~\cite{Satz:2001kf}. (See also the discussion in
Ref.~\cite{gliozzi}.) A similar estimate was given in
Ref.~\cite{Suzuki:2002sr}, based on the hypothesis of Abelian
dominance.

The conventional definition of the string breaking distance is the
distance $r_{sb}$, at
which the energy of two static-light mesons is equal to the energy of the
string, i.e.
\begin{equation}
 2m_{sl} = \sigma \cdot r_{sb}- \frac{\pi} {12r_{sb}}\,.
\end{equation}
The Wuppertal group found $r_{sb}=2.3 \, r_0$ at $m_\pi/m_\rho =
0.7$ \cite{Bolder:2001un}, while CP-PACS found $r_{sb}=2.2 \, r_0$
at $m_\pi/m_\rho=0.6$ \cite{Aoki:1998sb}. In full QCD it was found
$\sqrt{\sigma} r_0 = 1.14$~\cite{Bolder:2001un} and $\sqrt{\sigma}
r_0 = 1.16$~\cite{Allton:2002sk}, respectively, from which we
derive $2m_{sl} \approx 2.9/r_0 \approx 1.1~\mbox{GeV}$, assuming
$r_0=0.5$ fm. This agrees with the estimate of
\cite{Digal:2001iu}.
Using these values for $\sqrt{\sigma} r_0 $ and $m_{sl}$ and
assuming the mentioned above form of $C_E(r)$ we can estimate the
values of $r$ and $t$ at which the two terms in eq.(\ref{wl1})
become equal indicating that the string breaking effects become
visible:
\begin{equation}
r = t = \frac{4 m_{sl}}{\sigma}\,.
\end{equation}
We also estimate the numerical value of the Wilson loop of the
corresponding size:
\begin{equation}
\langle W(r,t) \rangle \lesssim 10^{-11} \cdot e^{-V_0 t}\,.
\end{equation}
It is a challenging task to record a Wilson loop of this order
of magnitude. Recently, a first successful attempt to do so was
reported in~\cite{Kratochvila:2003zj} where the authors studied
the adjoint static potential in three-dimensional SU(2)
gauge theory.

At finite temperature $T < T_c$ string breaking has been
studied in \cite{DeTar:1999qa}.
The heavy quark potential $V(r,T)$ is obtained from the Polyakov loop correlator:
\begin{equation}
\frac{1}{T} V(r,T)= - \mbox{ln} \langle L(\vec{s}) L^{\dagger}(\vec{s}\,')\rangle\,,
\label{hqp}
\end{equation}
up to an entropy contribution, where $r=|\vec{s} - \vec{s}\,'|$.
At large separations
\begin{equation}
\langle L(\vec{s}) L^{\dagger}(\vec{s}\,')\rangle
\underset{|\vec{s}-\vec{s}\,'| \rightarrow \infty}{\rightarrow} |\langle
L\rangle|^2\,,
\end{equation}
where $|\langle L\rangle|^2 \neq 0$, as global $Z_3$ is broken by
fermions. It should be noted that the potential in (\ref{hqp}) is
a color average, which is related to the proper singlet and the
octet potentials by~\cite{McLerran:pb,Nadkarni:1986cz}
\begin{equation}
e^{-V(r,T)/T} = \frac{1}{9} e^{-V_{sing}(r,T)/T} + \frac{8}{9}
e^{-V_{oct}(r,T)/T}\,.
\label{singlet}
\end{equation}
It would be desirable to compute the singlet potential introduced
in (\ref{singlet}). Work on calculating the singlet and octet
potential separately is in progress. In the recent publication
\cite{Kaczmarek:2003ph} this calculation was already performed for
$N_f=2$ lattice QCD on $16^3\, 4$ lattice.

The spectral representation of the Polyakov loop correlator is given by
\cite{Luscher:2002qv}
\beqn
\langle L(\vec{s}) L^{\dagger}(\vec{s}\,')\rangle  = \sum_{n=0}^{\infty}
w_n e^{-E_n(r)/T}\,,
\label{eq:series}
\eeqn
where $w_n$ are integers.
At zero temperature we have $V(r,0)=E_0(r)$, up to a constant, where $E_0$ is
the
ground state energy. At finite
temperature $V(r,T)$ gets contributions from all (excited) states.
As was discussed already, at $T=0$ the potential can be described by
the string model potential up to the string breaking distance $r_{sb}$.
Beyond this distance
the state of two static-light mesons becomes the ground state of the system.
Thus, there are two competing states in the spectrum, and it depends on the
distance $r$, which one will be the ground state. We may expect that the
situation at small temperature is similar to the case of $T=0$.

We shall now assume that at temperatures $T < T_c$ the Polyakov loop
correlator can be described in terms of these two states.
We then have
\beqn
\langle L(\vec{s}) L^{\dagger}(\vec{s}\,') \rangle =
e^{-(V_0(T)+V_{\mathrm{string}}(r,T))/T} + e^{-2E(T) /T}\,,
\label{eq:two:exp}
\eeqn
where the finite temperature string potential $V_{\rm string}(r,T)$ is given
by~\cite{Gao:kg}:
\beqn
V_{\mathrm{string}}(r,T) & = & -\frac{1}{r}
\Bigl(\alpha-\frac{1}{6} {\rm arctan}(2 r T)\Bigr)
\nonumber \\
& & + \Bigr(\sigma(T)+\frac{2 T^2}{3} {\mathrm{arctan}} \frac{1}{2r T}\Bigr)r +
\frac{T}{2} \ln \Bigl(1+4 r^2 T^2\Bigr)\,.
\label{eq:pot:string}
\eeqn
We consider the temperature dependent string tension $\sigma (T)$ as a free
parameter.
While in~\cite{Gao:kg} $\alpha$ was fixed at $\pi/12$, a fit of the
short distance part of the potential at $T=0$ gave~\cite{Bali:2000vr}
$\alpha = 0.32 \sim 0.34$. In the following we shall consider both cases,
$\alpha=\pi/12$ and $\alpha=0.33$.
The energy $E(T)$ can be written as
\beqn
 E(T)=\frac{1}{2}V_0 + m(T)\,,
\label{meff}
\eeqn
where $m(T)$ is the constituent quark mass \cite{Satz:2001kf}.

A long time ago the following ansatz for the Polyakov loop correlator
has been proposed~\cite{Karsch:1987pv}:
\beqn
\langle L(\vec{s}) L^{\dagger}(\vec{s}\,') \rangle =
e^{-(V_0(T)+V_{\mathrm{KMS}}(r,T))/T}\,,
\label{eq:one:exp}
\eeqn
where
\beqn
V_{KMS}(r,T)=\frac{\tilde{\sigma}}{\mu}(1-e^{-\mu r}) - \frac{\tilde\alpha}{r}
e^{-\mu r}\,.
\label{kms}
\eeqn
As we will see, this potential cannot capture the physics below the finite
temperature transition. We also do not consider this potential a valid ansatz,
because string breaking is a level crossing phenomenon.

Besides the non-Abelian potential, we will study the Abelian one. In
particular we shall be interested in its monopole and photon parts. From
studies at zero temperature~\cite{Bali:1996dm,Bornyakov:2001nd} it is known
that the monopole part of the potential decreases linearly down to very small
distances, showing no Coulomb term, which sometimes makes it easier to extract
a string tension. It appears that the monopole part of the potential has not
only no Coulomb term, but also shows no broadening of the flux tube as the
length of the flux tube is increased. (Both phenomena are connected of course.)
As we show below, our monopole potential is also linear at distances up to
the distance of order of 0.5~fm where flattening starts. Thus we may write
\begin{equation}
{\langle L_{\rm mon}(\vec{s}) L_{\rm mon}^\dagger(\vec{s}\,')\rangle}
= e^{-(V_0^{\rm mon}(T)+V_{\rm string}^{\rm mon}(r,T))/T} +
e^{-2E_{\rm mon}(T)/T}\,,
\label{mon_fit1}
\end{equation}
where
\begin{equation}
V_{\rm string}^{\rm mon}(r,T) = \sigma_{\rm mon} \cdot r\,,
\quad E_{\rm mon}(T)=V_0^{\rm mon}(T)+m_{\rm mon}(T)\,.
\label{mon_fit2}
\end{equation}

\subsection{Hypercubic blocking}
\label{IVB1}

As was mentioned already, we apply hypercubic blocking
(HCB)~\cite{Hasenfratz:2001hp} to reduce the statistical errors.
That means every SU(3) link matrix $U(s,\mu)$ is replaced by a new link
matrix $U_{HCB}(s,\mu)$, which is the weighted sum of products of link matrices
along paths from $s$ to $s+\hat{\mu}$ within adjacent cubes projected onto the
nearest SU(3) group element. We used the same parameters as
in~\cite{Hasenfratz:2001hp}.

\FIGURE[t]{
\centerline{\includegraphics[angle=-00,scale=0.45,clip=true]{hcb.eps}}
\vspace{-8mm}
\caption{Effect of hypercubic blocking on the potential at $\beta=5.2,
\kappa=0.1335$.}
\label{fig:hcb}
}
In Fig.~\ref{fig:hcb} we compare the static potential from blocked
and unblocked configurations. We see that the statistical errors
are substantially reduced. Furthermore, rotational invariance is
improved, in agreement with earlier
observations~\cite{Hasenfratz:2001hp}. The blocking procedure
decreases the self energy of the static sources, which causes a
constant shift in the potential. In Fig.~\ref{fig:hcb} we shift
the potential by $1.33/r_0$, so that it agrees with the unblocked
potential at $r=\sqrt{2}a$. We find good agreement at all
distances, except perhaps at $r=a$. The shift agrees with the
change in the asymptotic value of the potential, $-2T\log\,\langle
L\rangle$, which was found to be $1.24(25)/r_0$. The discrepancy
at $r=a$ can be accounted for by perturbative
corrections~\cite{Hasenfratz:2001tw}.
All our fits are made for $r/r_0 \ge 1$ thus this point is always discarded.

\subsection{Non-Abelian potential}

We first fit the static potential by the two-state ansatz
(\ref{eq:two:exp}). This is done for two different choices of $\alpha$,
$\alpha=\pi/12$ and $\alpha=0.33$. Examples of the fit for $T\slash T_c =
0.87$ and $0.98$ and the second choice $\alpha=0.33$ are shown in
Fig.~\ref{fig:potentials}. The curves for $\alpha=\pi/12$ and $\alpha=0.33$ are
practically indistinguishable from each other, visually and in terms of
$\chi^2/{\rm dof}$.
We also show the asymptotic value of the potential, $-2T\ln \langle  L\rangle$.
The potential converges to this value at large distances. The two-state ansatz
describes the data very well. The fit parameters
are given in Tables~\ref{tbl:fit:5.2}
and \ref{tbl:fit:5.25}, where $\sigma(0)=(1.14/r_0)^2$~\cite{Bolder:2001un}
was used.

\FIGURE[t]{
\includegraphics[angle=-00,scale=0.45,clip=true]{full.fits.eps}
\vspace{-3mm}
\caption{The heavy quark potential at $\beta=5.2$ for $T\slash T_c = 0.87$ and
  $0.98$, together with the fit using $\alpha=0.33$. The horizontal lines
  show the asymptotic value of the potential, where the shaded area indicates
  the error.}
\label{fig:potentials}
}
\TABLE[t]{
\begin{tabular}{|c|c|c|c|}
\hline
\multicolumn{4}{|c|}{$\alpha=0.33$}\\
\hline
$T/T_c$    & $V_0 r_0$ & $\sigma(T)/\sigma(0)$ & $ m(T) r_0 $ \\
\hline
~~~~0.80~~~~ &~~~2.33(2)~~~& ~~~0.86(3)~~~& 1.02(10)   \\
0.87        & 2.59(3) &  0.76(4) &  0.89(10)  \\
0.94        & 2.80(3) &  0.79(5) &  0.70(8)   \\
0.98        & 2.85(8) &  0.90(12)&  0.25(6)   \\
\hline
\hline
\multicolumn{4}{|c|}{$\alpha=\pi/12$}\\
\hline
0.80  & 2.00(2)  &  0.97(3)  &  1.19(11) \\
0.87  & 2.23(3)  &  0.89(4)  &  1.00(10)   \\
0.94  & 2.42(4)  &  0.95(5)  &  0.90(8)    \\
0.98  & 2.46(10) &  1.05(14) &  0.44(7)   \\
\hline
\hline
\multicolumn{4}{|c|}{Monopole part} \\
\hline
0.80 &~~~0.47(1)~~~& 0.90(1) & 1.24(4)     \\
0.87  &  0.53(1) & 0.85(1)  & 1.07(2)     \\
0.94  &  0.61(1) & 0.84(1) & 1.03(3)     \\
0.98  &  1.06(1) & 0.46(1) & 0.21(1)     \\
\hline
\end{tabular}
\vspace{-0mm}
\caption{Fit parameters of the two-state ansatz~\eq{eq:two:exp} at $\beta=5.2$, where we have
  assumed $\sigma(0)=(1.14/r_0)^2$.}
\label{tbl:fit:5.2}  }

\TABLE[t]{
\begin{tabular}{|c|c|c|c|}
\hline
\multicolumn{4}{|c|}{$\alpha=0.33$}\\
\hline
$T/T_c$    & $V_0 r_0$ & $\sigma(T)/\sigma(0)$ & $ m(T) r_0 $ \\
\hline
~~~~0.86~~~~   &~~~ 2.71(3)~~~ & ~~~0.74(5)~~~ & 0.81(7)   \\
0.92           & 2.87(7) &  0.70(10) &  0.37(7) \\
0.96           & 3.00(6) &  0.67(9)  &  0.25(7)  \\
0.97           & 3.12(7) &  0.57(10) &  0.15(7) \\
0.99           & 3.16(5) &  0.55(8)  &  0.28(7)  \\
\hline
\hline
\multicolumn{4}{|c|}{$\alpha=\pi/12$}\\
\hline
0.86   & 2.34(3) &  0.88(5)  &  1.00(6)  \\
0.92   & 2.49(7) &  0.86(10) &  0.57(7)   \\
0.96   & 2.61(6) &  0.83(10) &  0.45(7)   \\
0.97   & 2.71(6) &  0.73(9)  &  0.35(6)   \\
0.99   & 2.75(5) &  0.72(8)  &  0.48(7)   \\
\hline
\hline
\multicolumn{4}{|c|}{Monopole part}\\
\hline
0.86   & 0.56(2)  & 0.80(3)  & 1.18(8)  \\
0.92   & 0.57(1) & 0.80(2)   & 0.54(2)  \\
0.96   & 0.64(1)   & 0.78(2) & 0.33(1)  \\
0.97   & 0.66(1)  & 0.79(1)  & 0.27(1)  \\
0.99   & 0.65(1) & 0.79(1)   & 0.39(1)  \\
\hline
\end{tabular}
\vspace*{0.5cm}
\caption{Fit parameters of the two-state ansatz~\eq{eq:two:exp} at $\beta=5.25$, where we have
  assumed $\sigma(0)=(1.14/r_0)^2$.}
\label{tbl:fit:5.25}  }
\FIGURE{
\centerline{\includegraphics[angle=-00,scale=0.45,clip=true]{string.tension.na.eps}}
\vspace{-8mm}
\caption{The string tension from a fit of the two-state ansatz~\eq{eq:two:exp} as a function
  of temperature.
  The quenched value of
  the string tension \cite{Kaczmarek:2000mm} is shown for comparison. The shaded
  region indicate the error bar.}
\label{fig:sigma:mass:a}
}
\FIGURE{
\centerline{\includegraphics[angle=-00,scale=0.45,clip=true]{mass.na.eps}}
\vspace{-8mm}
\caption{The constituent quark mass from a fit
  of the two-state ansatz~\eq{eq:two:exp} as a function of temperature.
 The dash--dotted
  line indicate the zero--temperature quenched value of the mass.}
\label{fig:sigma:mass:b}
}

The string tension $\sigma(T)$ and the constituent quark mass $m(T)$ are
plotted in Figs.~\ref{fig:sigma:mass:a} and \ref{fig:sigma:mass:b}, respectively.
In Fig.~\ref{fig:sigma:mass:a}
the quenched value of $\sigma(T)/\sigma(0)$~\cite{Kaczmarek:2000mm} is shown
for comparison. Both the string tension and the
constituent quark mass decrease with increasing temperature, as we expect.
The results differ by approximately one $\sigma$ between $\alpha=0.33$ and
$\alpha=\pi/12$.
For lower temperatures, at $T/T_c=0.80$ and $T/T_c=0.87$, we
find rather good agreement between the results of the quenched
theory and our results obtained with the choice $\alpha=0.33$. For
higher temperatures agreement is much worse, especially for the
lighter quark mass. A reason for this discrepancy might be that
our Ansatz is not valid at temperatures close to the transition.

 Our values for the constituent quark mass $m(T)$ are larger
by about $100$ MeV than those reported in \cite{Digal:2001iu}.
However, one should note a difference between our definition
of the self energy and the definition used in
Ref.~\cite{Digal:2001iu}. 

We are now able to compute the string breaking distance $r_{sb}$
from
\begin{equation}
V_{\rm string}(r_{sb},T) = 2m(T)\,.
\end{equation}
In Figs.~\ref{fig:overlap:a} and \ref{fig:overlap:b} we show the two energy levels together with
the data. The string breaks where the two levels cross.
\FIGURE{
\centerline{\includegraphics[angle=-00,scale=0.45,clip=true]{levels.na.beta5.2kappa0.1335.eps}}
\vspace{-8mm}
\caption{The string potential and the constituent mass as a function of
distance at $\beta=5.2$ for $\kappa=0.1335$ ($T/T_c=0.87$). The shaded regions indicate the
errors.}
\label{fig:overlap:a}}
\FIGURE{
\centerline{\includegraphics[angle=-00,scale=0.45,clip=true]{levels.na.beta5.2kappa0.1343.eps}}
\vspace{-8mm}
\caption{The same as in Fig.~\ref{fig:overlap:a} but for
$\kappa=0.1343$ ($T/T_c=0.98$).}
\label{fig:overlap:b}}
The dependence of $r_{sb}$ on the temperature is shown in
Fig.~\ref{fig:rbr:na}. We see that $r_{sb}$ decreases as the temperature is
increased. The difference of $r_{sb}$ between the two choices of $\alpha$ lies
within the error bars.

\FIGURE{
\centerline{\includegraphics[angle=-00,scale=0.45,clip=true]{rbr.na.eps}}
\vspace{-8mm}
\caption{The string breaking distance $r_{sb}$ as a function of temperature.
The dash--dotted
  line indicate the corresponding zero--temperature quenched value.}
\label{fig:rbr:na}
}

Let us now consider the screening potential~(\ref{kms}). Fitting this potential
to our data gives a comparable value of $\chi^2/{\rm dof}$. However, the
parameters of the fit turn out to be unphysical. For example,
at $T\slash T_c=0.80$, $0.87$ and $0.94$ we obtain $\tilde{\sigma}/\sigma(0)=
21(6)$, $13(2)$ and $5.5(6)$, respectively. Only close to the deconfinement
transition we do find a reasonable value for the string tension:
$\tilde{\sigma}/\sigma(0) = 0.4(3)$ at $T/T_c = 0.98$.

The screening potential (\ref{kms}) may be rewritten (up to a constant) in the
following form \cite{Wong:2001uu}:
\begin{equation}
V^{\rm Wong}(r,T)=\left[ -\frac{4}{3} \frac{\alpha_s}{r} - \frac{b(T)}{\mu_0}
\right] e^{-\mu_0 r}\,.
\label{kms2}
\end{equation}
Taking (as in \cite{Wong:2001uu}) $b(T)=b_0 \left(1-(T/T_c)^2\right)$, $b_0=0.35$
GeV$\,^2$, $\mu_0= 0.28$ GeV, and $\alpha_s \sim 0.32$ ($\alpha_s \sim 0.24$)
for the charmonium (bottonium) potential, and shifting the potential by a
constant so that it agrees with the lattice potential at $r=r_0$, we find no
agreement between this potential and the lattice data. Thus, the quarkonium
spectra derived from this potential~\cite{Wong:2001uu} need to be revised.

\subsection{Monopole part of the potential}

We carried out a similar analysis as before for the monopole part of the
heavy quark potential, which is obtained from the correlator (\ref{mon_fit1}).
The fit parameters are given in Tables~\ref{tbl:fit:5.2} and
\ref{tbl:fit:5.25}, and the potential is shown in
Fig.~\ref{fig:mon:potentials}.  The errors are smaller than in the previous
fits, as expected. The monopole part of the potential shows no Coulomb term,
while at large distances it converges to its
asymptotic value $-2T\ln \langle L_{\mathrm{mon}}\rangle$.

\FIGURE{
\centerline{\includegraphics[angle=-00,scale=0.45,clip=true]{monopole.fits.eps}}
\vspace{-8mm}
\caption{The monopole part of the potential as a function of distance at
  $\beta=5.2$ for $T\slash T_c = 0.87$ and $0.98$, together with a fit
  of the form (\ref{mon_fit2}) (dashed curve). The horizontal lines show the
  asymptotic value of the potential, where the shaded area indicates the
  error.}
\label{fig:mon:potentials}
}

\FIGURE{
\centerline{\includegraphics[angle=-00,scale=0.45,clip=true]{string.tension.mon.eps}}
\vspace{-8mm}
\caption{The string tension of the
  monopole part of the potential as a function of temperature.
  In the left figure the quenched value of
  the string tension~\cite{Kaczmarek:2000mm} is shown for comparison. The shaded
  region indicate the error bar.}
\label{fig:sigma:mon:a}
}
\FIGURE{
\centerline{\includegraphics[angle=-00,scale=0.45,clip=true]{mass.mon.eps}}
\vspace{-8mm}
\caption{The same as in Fig.~\ref{fig:sigma:mon:a} but for the constituent quark mass.
The dash--dotted line indicate the zero--temperature quenched value of the mass.}
\label{fig:sigma:mon:b}
}

The string tension $\sigma_{\rm mon}(T)$ and the constituent quark mass
$m_{\rm mon}(T)$ are shown in Fig.~\ref{fig:sigma:mon:a} and \ref{fig:sigma:mon:b},
respectively.
Because the Coulomb term is absent, we now can determine the string tension
much more accurately. We find substantially larger values than in the quenched
case. Furthermore, the string tension appears to decrease more slowly as the
the system is heated. The constituent quark mass $m_{\rm mon}(T)$
looks very much the same as in the non-Abelian case. The same
holds for the string breaking distance, $r^{\rm mon}_{sb}$, which
is shown in Fig.~\ref{fig:rbr:mon}.

\FIGURE{
\centerline{\includegraphics[angle=-00,scale=0.45,clip=true]{rbr.mon.eps}}
\vspace{-8mm}
\caption{The string breaking distance $r^{\rm mon}_{sb}$ obtained from the
monopole potential as a function of temperature.
 The dash--dotted
  line indicate the corresponding zero--temperature quenched value.}
\label{fig:rbr:mon}
}

To shed further light on the string breaking mechanism, we have computed
the action density, the color electric field and the monopole current in the
vicinity of the (broken) string.

The definitions of observables are the same as in Ref.~\cite{Bornyakov:2001nd}.
We are interested in local Abelian operators of the form:
\beqn
\cO(s) =
\mbox{diag} (\cO_1(s),\cO_2(s),\cO_3(s)) \in U(1)\times U(1)\,.
\eeqn
The correlator of the action density -- which is $C$-parity even operator --
with the product of the monopole Polyakov loops,
$\cL_{\rm mon}(\vec{s'}) \cL_{\rm mon}^\dagger(\vec{s}\,'')$,
can be written analogously to Ref.~\cite{Bali:1994de}:
\begin{equation}
\langle  \cO(s) \rangle_{\cL_{\rm mon}} \equiv \frac{1}{3}
\frac{\langle\mbox{Tr}\,{\cO}(s) \mbox{Tr}\, [\cL_{\rm mon}(\vec{s'}) \cL_{\rm mon}^\dagger(\vec{s}\,'')]
\rangle}{\langle \mbox{Tr}\, [\cL_{\rm mon}(\vec{s'}) \cL_{\rm mon}^\dagger(\vec{s}\,'')]\rangle}
- \frac{1}{3} \langle \mbox{Tr} {\mathcal
O}\rangle \,,\label{eq:operator_even}
\end{equation}
where
\beqn
\cL_{\rm mon}(s) =
\mbox{diag} (L^{\rm mon}_1(s),L^{\rm mon}_2(s),L^{\rm mon}_3(s))\,,
\eeqn
($cf.$, Eq.~\eq{eq:monpl}).

As for the $C$-parity odd operators $\cO$, such as the color electric field
and the monopole current, we have
\begin{equation}
\langle \cO(s)  \rangle_{\cL_{\rm mon}}  \equiv \frac{\langle\mbox{Tr}\,({\cO}(s)\,
[\cL_{\rm mon}(\vec{s'}) \cL_{\rm mon}^\dagger(\vec{s}\,'')])\rangle}{\langle
\mbox{Tr}\, [\cL_{\rm mon}(\vec{s'}) \cL_{\rm mon}^\dagger(\vec{s}\,'')] \rangle}\,,
\label{eq:operator_odd}
\end{equation}
in analogy to the case of $SU(2)$ and $U(1)$ theories, 
Refs.~\cite{Haym93}.

The monopole part of the action density $\rho_A^{\cL_{\rm mon}}$, the monopole part
of the color electric field $E_i^{\cL_{\rm mon}}$ and the
monopole current $k^{\cL_{\rm mon}}$, induced by the Polyakov loops, are then given by
\begin{equation}
\rho_A^{\cL_{\rm mon}}(s) = \frac{\beta}{3} \sum_{\mu>\nu} \langle
\mbox{diag} (\cos(\theta_1^{\rm mon}(s,\mu,\nu)),\cos(\theta_2^{\rm mon}(s,\mu,\nu)),
\cos(\theta_3^{\rm mon}(s,\mu,\nu)))\rangle_{\cL_{\rm mon}} \,,
\end{equation}
where the plaquette angles, $\theta^{\rm mon}_i(s,\mu,\nu)$, are constructed from the
monopole link angles~\eq{eq:theta:mon},
\beqn
\label{eq:def-elf}
E_j^{\cL_{\rm mon}}(s) = {\rm i}\, \langle \mbox{diag} (\theta_1^{\rm mon}(s,4,j),
\theta_2^{\rm mon}(s,4,j),\theta_3^{\rm mon}(s,4,j))\rangle_{\cL_{\rm mon}}\,,
\eeqn
and
\beqn
\label{eq:def-cur}
k^\cL(^*s,\mu) = 2 \pi {\rm i}\, \langle \mbox{diag}
(k_1(^*s,\mu),k_2(^*s,\mu),k_3(^*s,\mu)) \rangle_{\cL_{\rm mon}}\,,
\eeqn
respectively. 

\begin{figure}[!htb]
\begin{center}
\hspace{-4cm}\epsfig{file=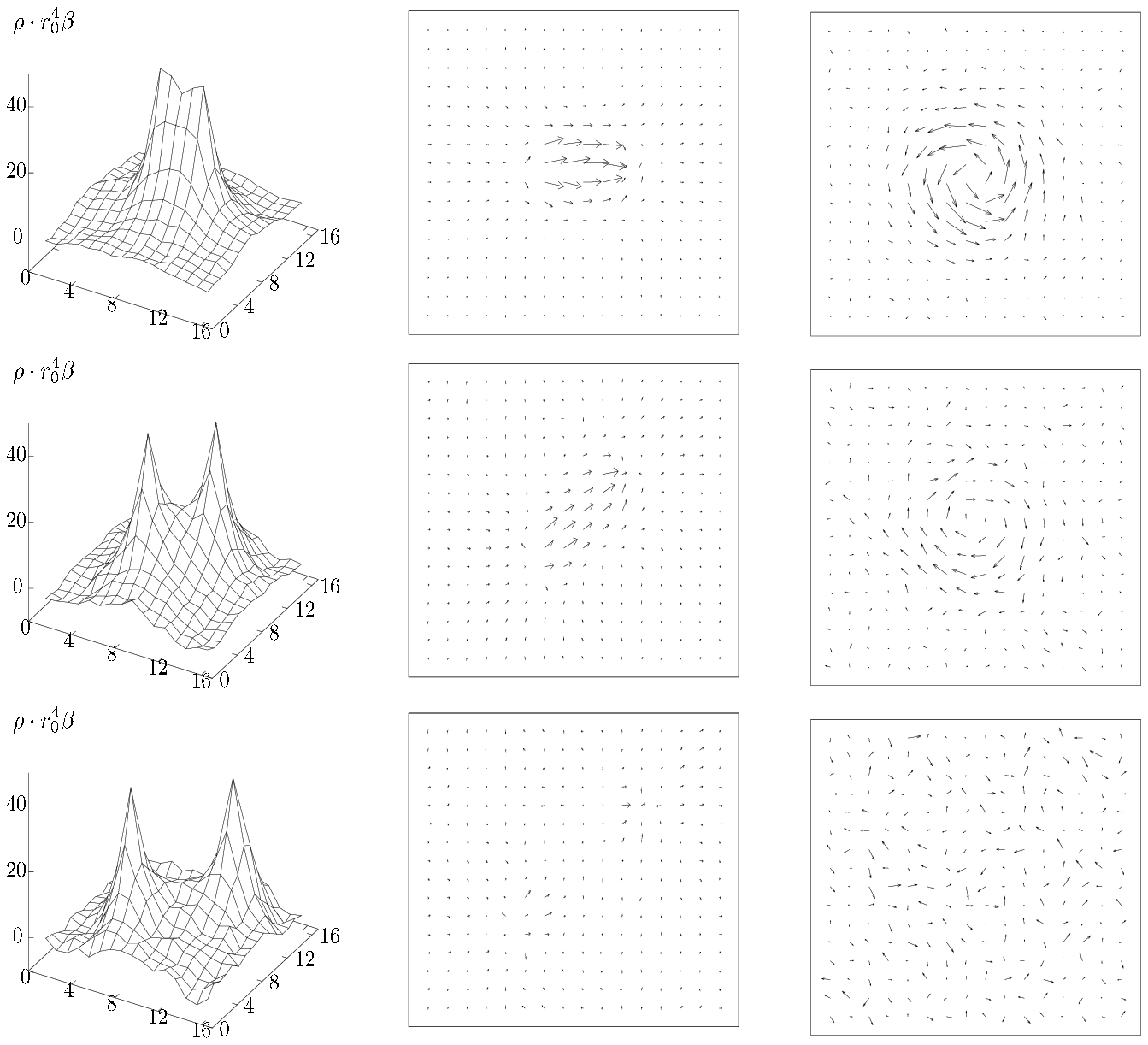,width=17cm,clip=}
\vspace{-14cm}
\hspace{4cm}(a) \hspace{4cm}(b) \hspace{4cm}(c)
\end{center}
\caption{The monopole part of the action density (a), the monopole part of the
color electric field (b) and the solenoidal monopole current in the plane
perpendicular to the flux tube (c) at $T/T_c=0.98$ and
distances (from top to bottom) 0.5, 0.8 and 1.3 fm.}
\label{fig:act_dens}
\end{figure}

In Fig.~\ref{fig:act_dens} we show the result for $T/T_c=0.98$ and three
different separations, $r=0.5$, 0.8 and 1.3 fm. Our estimate of the string
breaking distance at this temperature is $\approx 0.5$ fm.
The figure suggest that the flux tube has disappeared at the latest at $r=1.3$
fm.

\section{Monopole density}

Another characteristic quantity of the confining vacuum is the monopole
density, which we define as
\beqn
\rho = \frac{1}{12 N_t N_s^3} \Bigl \langle \sum_{i=1}^3\sum_{s,\mu}
|k_{i}(^*s,\mu)|
\Bigr\rangle\,,
\eeqn
where the monopole current, $k_{i}(^*s,\mu)$, is given in
(\ref{eq:mon:current}).

\FIGURE{
\centerline{\includegraphics[angle=-00,scale=0.45,clip=true]{monopole.density.eps}}
\vspace{-8mm}
\caption{The monopole density at $\beta = 5.2$ as a function of temperature.}
\label{fig:dens:a}
}

In Fig.~\ref{fig:dens:a} we compare the monopole density of this work
with that of the quenched theory. The quenched result has been obtained on the
same sized lattice at $\beta=5.8$, 5.9, 6.0, 6.1 and 6.2.
The density in full QCD is substantially higher than in the quenched
theory, in agreement with our earlier result at $T=0$~\cite{Bornyakov:2001nd}.
We believe that the introduction of dynamical fermions causes an attraction
between monopoles and antimonopoles, which naturally leads to an increase in
the monopole density. A similar mechanism has been observed in the case of
instantons and antiinstantons~\cite{Hasenfratz:1999ng}. Both mechanisms are,
of course, related, because (anti-)instantons are intimately connected with
monopoles~\cite{Hart:1995wk}.

Near the finite temperature transition we expect the monopoles to
gradually become static as the temperature becomes high.
This can be monitored by the
asymmetry of the density of spatial and temporal monopole
currents~\cite{Brandstater:1991sn,kitahara}:
\beqn
\eta = \frac{\rho_t-\rho_s}{\rho_t+\rho_s}\,,
\label{eq:eta}
\eeqn
where $\rho_t$($\rho_s$) is the density of the temporal (spatial)
monopole currents,
\beqn
\rho_t = \frac{1}{3 N_t N_s^3} \Bigr\langle \sum_{i=1}^3\sum_{s} |k_{i}(^*s,4)|  \Bigr\rangle\,,
\quad
\rho_s = \frac{1}{9 N_t N_s^3} \Bigr\langle \sum_{i=1}^3\sum_{s}\sum^3_{\mu=1} |k_{i}(^*s,\mu)|
\Bigr\rangle\,.
\eeqn
If all currents are time-like, then this quantity is unity, while in the case
of an isotropic distribution it is zero.
\FIGURE{
\centerline{\includegraphics[angle=-00,scale=0.45,clip=true]{monopole.density.asymmetry5.eps}}
\vspace{-8mm}
\caption{The asymmetry of the monopole
density at $\beta = 5.2$ as a function of temperature.}
\label{fig:dens:b}
}
In Fig.~\ref{fig:dens:b} we plot the asymmetry $\eta$ as a function of
temperature. We compare the result with the predictions of the quenched
theory. It is found that $\eta$ is zero in the confined phase and nonzero
in the deconfined phase.
In the deconfinement phase the value of $\eta$ is about 5
times smaller in full QCD compared to the quenched theory. A
reason for this may be rooted in a different nature of the
transition which is of the first order phase in the quenched case
while in full QCD one observes a smooth crossover.

\section{Conclusions}

We have studied QCD with two flavors of dynamical quarks at finite
temperature on a $16^3\,8$ lattice. At the phase transition the lattice
spacing is $a \approx 0.12$ fm. We employed non-perturbatively improved Wilson
fermions, so that we may expect finite cut-off effects to be small.

To determine the parameters of the transition, notably the transition
temperature and the string tension, and to shed light on the
dynamics of the transition, it helped to resort to Abelian variables in the
maximally Abelian gauge.

We observed string breaking in Polyakov loop correlators. This is a level
crossing phenomenon. Accordingly, we fitted the correlator by a two-state
ansatz, consisting of a string state and a two-meson state. We found good
agreement of this ansatz with our numerical data for $T \lesssim T_c$, while
we could rule out previously proposed single-state correlation functions.
The string breaking distance was found to be $r_{\rm sb} \approx 1$ fm at
$T/T_c \approx 0.8$, our lowest temperature. String breaking is also clearly
visible in the action density, the color electric field distribution and the
solenoidal monopole current around the static sources.

To make contact with the chiral limit, we need to increase $N_t$, because
the coupling cannot be taken smaller than $\beta \approx 5.2$. Work on
$24^3\,10$ lattices at $m_\pi/m_\rho \approx 0.6$ is in progress. Preliminary
results have been presented in~\cite{Nakamura}.

\begin{acknowledgments}
We like to thank Alan Irving and Dirk Pleiter for assistance. The calculations
have been performed on the Hitachi SR8000 at KEK Tsukuba and on the MVS 1000M
at Moscow. We like to thank the staff of the Moscow Joint Supercomputer
Center, especially A.V. Zabrodin, for their support. We furthermore thank
Ph. De Forcrand, V. Mitrjushkin and M. M\"uller-Preussker for useful
discussions.
This work is partially supported by grants INTAS-00-00111,
RFBR 02-02-17308, RFBR 01-02-17456, RFBR-DFG 03-02-0416, RFBR 03-02-16941,
DFG-RFBR 436RUS113/739/0 and CRDF award RPI-2364-MO-02.
M.Ch. is supported by JSPS Fellowship P01023. V.B. acknowledges
support from JSPS RC30126103. T.S. is
supported by JSPS Grant-in-Aid for Scientific Research on Priority Areas
13135210 and 15340073.
\end{acknowledgments}

\end{document}